\documentclass[aps,prb,amsmath]{revtex4}
\usepackage{graphicx}% Include figure files
\usepackage{bm}% bold math
\usepackage{bbold}

\def\br{ \bm{r} }

\def\bq{ \bm{q} }
\def\bnab{ \bm{\nabla} }
\def\bfn{ \bm{n} }
\def\bA{ \bm{A} }

\def\bD{ \bm{D} }

\def\hby{ \hat{\bm{y}} }
\def\bjs{ \bm{j}_s }
\def\im{ \mathrm{Im}\, }
\def\re{ \mathrm{Re}\, }

\begin{document}
\title{Fulde-Ferrell-Larkin-Ovchinnikov superconductors near a surface}

\author{K. V. Samokhin and B. P. Truong}

\affiliation{Department of Physics, Brock University, St. Catharines, Ontario L2S 3A1, Canada}
\date{\today}

\begin{abstract}
We show that the behaviour of the Fulde-Ferrell-Larkin-Ovchinnikov (FFLO) superconductors near a surface is considerably different from the usual case. The order parameter of the FF state is strongly deformed 
near the surface, which leads to a number of unusual features in the linear magnetic response, such as ``anti-screening'' or ``over-screening'' of the applied field. In a fully isotropic FF case,
the Meissner effect is still present, despite the vanishing of the transverse superfluid density in the bulk. 
We also calculate the surface critical field $H_{c3}$, which exhibits a peculiar temperature dependence.     
\end{abstract}

\maketitle

\section{Introduction}
\label{sec: Intro}

It was shown by Fulde and Ferrell\cite{FF64} (FF) and Larkin and Ovchinnikov\cite{LO64} (LO) that by increasing the spin splitting of the electron bands in a singlet superconductor 
one can drive the system into a peculiar nonuniform superconducting state, known as the FFLO state.
Due to the Cooper pairs in an FFLO superconductor having a nonzero center-of-mass momentum, the order parameter is periodically modulated, for instance, $\psi(\br)\propto e^{i\bm{q}\br}$, which corresponds to the 
FF state. More complicated structures containing two or more plane waves, such as the LO state with $\psi(\br)\propto\cos\bm{q}\br$, are also possible.   

Experimental realization of the FFLO superconductivity has remained a challenge, because it requires a weak orbital pair breaking and a sufficiently clean sample. For these reasons, the search for the FFLO state has focused on
materials with low effective dimensionality and unconventional pairing, see Ref. \onlinecite{FFLO-review-07} for a review. A particularly strong evidence of the FFLO state has been recently found in quasi-two-dimensional (quasi-2D) 
organic compounds, such as $\lambda$-(BETS)$_2$FeCl$_4$ (Ref. \onlinecite{Uji12}). 
While proposed originally in the context of solid-state superconductors, the FFLO states are now recognized as a universal feature of paired fermionic systems with mismatched Fermi surfaces, ranging from
``cold'' Fermi gases\cite{RS09} to color-superconducting quark matter.\cite{CasNar04} 

In this paper, we aim to resolve a long-standing puzzle about the Meissner effect in the FFLO state. It has been known since the seminal work of Fulde and Ferrell\cite{FF64} that the transverse (relative to $\bq$) 
components of the superfluid density tensor, or the transverse phase stiffness, 
vanish in the isotropic FF state, which seemingly implies the absence of magnetic field expulsion from the superconductor in certain geometries. It is important to realize, however, that the Meissner effect measurements 
are done in finite samples, therefore the conclusions based on the expressions for the superfluid density tensor in the bulk are not necessarily valid in the presence of a boundary. To properly calculate the magnetic
response in a realistic situation, we derive the boundary conditions for the order parameter in a half-infinite FFLO superconductor and show that the superconducting state is necessarily deformed near the surface. 
As a result, the Meissner effect in the FF state is still present, albeit in a much changed form.    

Our second goal is the calculation of the surface critical field in an FFLO superconductor. If an external magnetic field $H$ is applied parallel to a superconductor-insulator interface, 
then in the usual (non-FFLO) case superconductivity first nucleates at $H=H_{c3}$, which is higher than the upper critical field $H_{c2}$ by a universal factor: 
$H_{c3}(T)\simeq 1.695H_{c2}(T)$ (Ref. \onlinecite{DeGennes-book}), with both critical fields having a linear temperature dependence. 
At $H_{c2}<H<H_{c3}$, the order parameter is localized near the surface. One can expect that in the FFLO superconductors this effect would be very different, if present at all,
just by looking at the temperature dependence of the FFLO upper critical field. In contrast to the usual case, $H_{c2}$ exhibits the ``Little-Parks'' oscillations,
due to the contribution of the Cooper pairs in the higher Landau levels.\cite{BK84} In quasi-2D materials, the orbital pair breaking effects can be probed by tilting 
the external field out of the basal plane, which results in the oscillations of the superconducting critical temperature as a function of the tilt angle.\cite{tilted-H}

Regarding the methodology, it is convenient to study the magnetic properties of the FFLO superconductors using a modified Ginzburg-Landau (GL) formalism. If the coefficient in front of the quadratic gradient term $|\bnab\psi|^2$ 
in the GL free energy is negative, but the quartic gradient term $|\bnab^2\psi|^2$ is positive, then the preferred superconducting state is modulated with a nonzero wavevector. Microscopic derivation in the simplest model of 
a clean paramagnetically-limited isotropic superconductor shows that the quadratic gradient term indeed changes sign in a sufficiently strong magnetic field, but it does so simultaneously with the coefficient in front 
of the $|\psi|^4$ term. Therefore, in order to ensure stability, one has to include higher-order terms, such as $|\psi|^6$, and others.\cite{BK97}
Then, the most stable state corresponds to a nonlinear generalization of the LO state, with the gap magnitude periodically modulated in space. On the other hand, in the presence of 
disorder and the pairing anisotropy, the $|\psi|^4$ term can remain positive while the $|\bnab\psi|^2$ term changes sign, thus stabilizing the simple single-plane wave FF state,
which is separated from the normal state by a second-order phase transition.\cite{AY01,HM06}

The paper is organized as follows. In Sec. \ref{sec: GL}, we introduce the modified GL description of the FFLO superconductors and discuss the issues with the Meissner effect in the FF state. 
In Sec. \ref{sec: Meissner}, we find the order parameter texture and the magnetic field distribution in a half-infinite FF superconductor. In Sec. \ref{sec: Hc3}, the surface critical field is calculated.  
Sec. \ref{sec: Conclusions} concludes with a summary of our results. Throughout the paper, $e$ denotes the absolute value of the electron charge.

\section{Ginzburg-Landau description of the FFLO states}
\label{sec: GL}

We consider a quasi-2D spin-singlet superconductor with the $xy$ plane being the basal plane and the order parameter depending only on $\br=(x,y)$. 
In order to split the electron bands and drive the system into the FFLO regime at low temperatures, 
we apply a sufficiently strong uniform magnetic field $\bm{H}_\parallel$ parallel to the plane.
The orbital effects, in particular, the Meissner effect and the surface critical field, are probed by tilting the magnetic field out of the plane, so that $H_z\neq 0$ and $B_z(x,y)=\nabla_xA_y-\nabla_yA_x\neq 0$. Here 
$\bA=(A_x,A_y)$ is the ``orbital'' vector potential, and we use the gauge $\bnab\cdot\bA=0$. 
Below the ``zero field'' limit always refers to the situation when the orbital effects are absent, \textit{i.e.} $H_z=B_z=0$, and $A_x=A_y=0$, but $\bm{H}_\parallel\neq\bm{0}$. 

Formation of a nonuniform superconducting state can be described phenomenologically by a modified GL functional in 2D, ${\cal F}=\int F d^2\br$, with the free energy density given by
\begin{equation}
\label{GL-energy-density-anisotropic}
  F = \alpha |\psi|^2 + \frac{\beta}{2}|\psi|^4 +K|\bD\psi|^2 + \tilde{K}|\bD^2\psi|^2+\varepsilon\tilde K\left(|D_x^2\psi|^2+|D_y^2\psi|^2\right).
\end{equation}
Here $\bD=\bnab+i(2e/\hbar c)\bA$ is the covariant derivative and $\alpha=a(T-T_{c,0})$, with $T_{c,0}$ being the critical temperature of the transition into a uniform superconducting state. 
In order for the instability with a finite wavevector to occur, we put $K<0$ but $\tilde K>0$. According to the discussion in the Introduction, we assume that $\beta>0$,  
therefore the bulk state is of the FF form and the superconductor-normal transition is of the second order.
The additional gradient terms, with $\varepsilon>-1$ being a dimensionless parameter, are included to describe the effects of the in-plane crystal anisotropy, appropriate for a 2D square lattice.\cite{BMS07} 

If $-1<\varepsilon<0$, then the bulk superconducting state in zero field is modulated along one of the principal axes, for instance,  
\begin{equation}
\label{bulk-FF-anisotropic}
  \psi(\br)=\Delta_0 e^{\pm iq_0x},
\end{equation}
where 
\begin{equation}
\label{FF-bulk-wavevector}
  q_0=\sqrt{\frac{|K|}{2(1+\varepsilon)\tilde K}}
\end{equation}
and $\Delta_0=\sqrt{a(T_c-T)/\beta}$. This solution exists below the critical temperature
\begin{equation}
\label{T_c-FF}
  T_c=T_{c,0}+\frac{K^2}{4a(1+\varepsilon)\tilde K},
\end{equation}
which is higher than that of a uniform state.
If $\varepsilon>0$, then at zero field the bulk FF state is modulated along one of the diagonals of the square lattice, \textit{\textit{i.e.}} $\psi(\br)\propto e^{i\bq\br}$, with $q_x^2=q_y^2=|K|/2(2+\varepsilon)\tilde K$.

In order to highlight the issues with the Meissner effect in the FF state, let us consider the isotropic case, \textit{i.e.} put $\varepsilon=0$ in Eq. (\ref{GL-energy-density-anisotropic}).
At zero field, we obtain from Eq. (\ref{GL-energy-density-anisotropic}) the nonlinear GL equation
\begin{equation}
\label{GL-eq-zero-field}
  \alpha\psi+\beta|\psi|^2\psi-K\nabla^2\psi+\tilde K\nabla^4\psi=0,
\end{equation}
which has an isotropically degenerate solution of the form
\begin{equation}
\label{isotropic-FF-state}
  \psi(\br)=\Delta_0e^{i\bq\br},
\end{equation}
where the optimal wavevector is given by $|\bq|=q_0=\sqrt{|K|/2\tilde K}$.
The supercurrent $\bjs=-c(\delta{\cal F}/\delta\bA)$ can be expanded in powers of the vector potential: $\bjs=\bjs^{(0)}+\bjs^{(1)}+{\cal O}(A^2)$, where 
\begin{equation}
\label{j_0}
  \bjs^{(0)}=-\frac{4e}{\hbar} \im \left\{ K \psi^* \bnab\psi + \tilde{K}\left[ (\bnab\psi)^*\nabla^2\psi-\psi^*\bnab\nabla^2\psi \right] \right\}
\end{equation}
is the spontaneous supercurrent, while the linear response to $\bA$ is given by 
\begin{equation}
\label{j_1}
  j_{s,i}^{(1)}=-\frac{8e^2}{\hbar^2c} \re \left\{ K|\psi|^2A_i-2\tilde K\left[(\psi^*\nabla^2\psi)A_i+\nabla_i(\psi^*\bA\bnab\psi)-2(\nabla_i\psi^*)\bA\bnab\psi\right]\right\}.
\end{equation}
Note the difference of Eqs. (\ref{j_0}) and (\ref{j_1}) from the textbook expressions.\cite{Tinkham-book} The presence of the higher-order gradient terms and the fact that $K<0$ dramatically change 
the way the FFLO superconductors conduct electric current.\cite{ST17}

It is easy to see from Eqs. (\ref{j_0}) and (\ref{j_1}) that the FF state (\ref{isotropic-FF-state}) carries no spontaneous current and its response to a weak external field is given by
\begin{equation}
\label{longitudinal-response}
  \bjs^{(1)}=-\frac{32e^2}{\hbar^2c}\tilde K\Delta_0^2\bq(\bq\bA).
\end{equation}
Thus we have reproduced the well-known observation\cite{FF64} that there is no linear response of the FF state in a fully isotropic system to a vector potential which is transverse to $\bq$, 
see also Ref. \onlinecite{zero-stiffness}. This surprising property actually holds beyond the GL model. Indeed, since the total free energy of the isotropic FF state depends only on $|\bq|$, 
the gauge invariance dictates that a uniform vector potential enters $F$ only via $|\bq-\bA|$. The supercurrent can be written as $j_{s,i}^{(1)}=-Q_{ij}A_j$, where
$$
  Q_{ij}=c\left.\frac{\partial^2F(|\bq-\bA|)}{\partial A_i\partial A_j}\right|_{\bA=\bm{0}}=cF''(q)\hat q_i\hat q_j+cF'(q)\frac{\delta_{ij}-\hat q_i\hat q_j}{q}.
$$
The second term here vanishes, because the equilibrium FF state with $q=|\bq|=q_0$ corresponds to the minimum of the free energy, and one obtains $Q_{ij}\propto\hat q_i\hat q_j$, \textit{i.e.} a purely longitudinal response. 

We would like to stress that the result (\ref{longitudinal-response}) should not be taken as a proof of the absence of the transverse current response and the Meissner effect in a realistic FF superconductor. 
The point is that the Meissner effect is measured in a finite sample and the applicability of Eq. (\ref{longitudinal-response}) in the presence of a surface is questionable. 
In fact, we will show below that the simple single-plane wave solution of the form 
(\ref{isotropic-FF-state}) does not satisfy the boundary conditions for the GL equations, leading to a considerable modification of the FF state near a surface and the restoration of the Meissner effect.

\section{Meissner effect}
\label{sec: Meissner}

In this section, we consider a half-infinite superconductor, with a straight surface at $x=0$. The GL free energy has the general anisotropic form given by Eq. (\ref{GL-energy-density-anisotropic}). 
We assume that $-1<\varepsilon\leq 0$ and that the zero-field order parameter depends only on $x$: $\psi(\br)=\psi(x)$. The response of a given order parameter texture to a weak external field 
perpendicular to the plane is obtained by solving the Maxwell equation 
\begin{equation}
\label{Maxwell-eq}
  \nabla^2\bA=-\frac{4\pi}{c}\bjs, 
\end{equation}
where the supercurrent is calculated in the linear approximation in the vector potential. In the Landau gauge, $\bA=A(x)\hby$, the ``orbital'' magnetic induction is given by $B_z=B(x)=dA/dx$.
 It follows from Eq. (\ref{GL-energy-density-anisotropic}) that the spontaneous supercurrent has the form 
\begin{equation}
\label{j_0-anisotropic}
  j_{s,x}^{(0)}=\frac{4e}{\hbar} \im \left[ |K|\psi^* \nabla_x\psi + (1+\varepsilon)\tilde{K}\left(\psi^*\nabla_x^3\psi-\nabla_x\psi^*\nabla_x^2\psi\right)\right],
\end{equation}
while $j_{s,y}^{(0)}=0$. For the linear response term we obtain:
\begin{equation}
\label{j_1-anisotropic}
  j_{s,y}^{(1)}=\frac{8e^2}{\hbar^2c} \re \bigl( |K||\psi|^2+2\tilde K\psi^*\nabla_x^2\psi\bigr) A
\end{equation}
and  $j_{s,x}^{(1)}=0$.

\subsection{Order parameter texture}
\label{sec: half-DW}

In order to find the zero-field order parameter in a half-infinite sample one has to solve the GL equation 
\begin{equation}
\label{GL-eq-anisotropic}
  \alpha\psi+\beta|\psi|^2\psi+|K|\nabla_x^2\psi+(1+\varepsilon)\tilde K\nabla_x^4\psi=0,
\end{equation}
supplemented by some boundary conditions for the order parameter at $x=0$. We propose that the first boundary condition is
\begin{equation}
\label{zero-gradient}
  \left.\nabla_x\psi\right|_{x=0}=0.
\end{equation}
This form was shown in Ref. \onlinecite{DeGennes-book} to be appropriate for an interface between a non-FFLO superconductor and an insulator.
Since Eq. (\ref{zero-gradient}) is not sensitive to the signs of the coefficients in the GL gradient terms, it is reasonable to apply it in the FFLO case as well. However, in contrast to the non-FFLO case, 
the condition (\ref{zero-gradient}) alone does not guarantee that the supercurrent normal to the surface vanishes, due to the presence of higher-order gradient terms, see Eq. (\ref{j_0-anisotropic}). 
Therefore, we have to additionally require that
\begin{equation}
\label{zero-normal-current}
  \left.j_{s,x}\right|_{x=0}=0.
\end{equation}
A more formal justification of the boundary conditions (\ref{zero-gradient}) and (\ref{zero-normal-current}) is presented in the Appendix.

To solve the nonlinear equation (\ref{GL-eq-anisotropic}), it is convenient to use the amplitude-phase representation of the order parameter: $\psi(x)=\Delta(x)e^{i\theta(x)}$. We obtain:
$$
  \alpha\Delta+\beta\Delta^3+|K|(\nabla_x+iv_s)^2\Delta+(1+\varepsilon)\tilde K(\nabla_x+iv_s)^4\Delta=0,
$$
where $v_s=\nabla_x\theta$ can be called the superfluid velocity, by analogy with its non-FFLO counterpart.\cite{Tinkham-book} The real and imaginary parts of the last equation are given by
\begin{equation}
\label{re-part-GL-eq}
  \alpha\Delta+\beta\Delta^3+|K|\hat R_1\Delta+(1+\varepsilon)\tilde K(\hat R_1^2-\hat R_2^2)\Delta=0
\end{equation}
and
\begin{equation}
\label{im-part-GL-eq}
  |K|\hat R_2\Delta+(1+\varepsilon)\tilde K\{\hat R_1,\hat R_2\}\Delta=0,
\end{equation}
respectively. Here $\hat R_1=\nabla_x^2-v_s^2$, $\hat R_2=\{\nabla_x,v_s\}$, and the curly brackets denote the anticommutator of two operators. It follows from Eq. (\ref{j_0-anisotropic}) that the supercurrent is given by 
$j_{s,x}^{(0)}=(4e/\hbar)I$, where
\begin{equation}
\label{conserved-I}
  I=|K|\Delta^2v_s+(1+\varepsilon)\tilde K\left[\Delta^2\nabla_x^2v_s-2\Delta^2v_s^3+4\Delta(\nabla_x^2\Delta)v_s+2\Delta(\nabla_x\Delta)(\nabla_xv_s)-2(\nabla_x\Delta)^2v_s\right].
\end{equation}
It is straightforward to check that $\nabla_x I$ is equal to the left-hand side of
Eq. (\ref{im-part-GL-eq}) multiplied by $\Delta$, which yields the current conservation condition: 
$$
  I(x)=\mathrm{const},
$$ 
at all $x$. In the absence of an external current injected into the system, we have $I=0$.

In an infinite system, Eq. (\ref{re-part-GL-eq}) supplemented by the condition $I=0$ has a trivial solution $\Delta(x)=\Delta_0$, $v_s^2(x)=|K|/2(1+\varepsilon)\tilde K=q_0^2$, see Eq. (\ref{FF-bulk-wavevector}), 
which corresponds to the single-plane wave FF state (\ref{bulk-FF-anisotropic}). 
However, it is easy to see that although this state carries zero current, it does not satisfy the boundary condition (\ref{zero-gradient}) and therefore will be inevitably modified near the surface. 

\begin{figure}
    \centering
  	\includegraphics[width=0.60\linewidth]{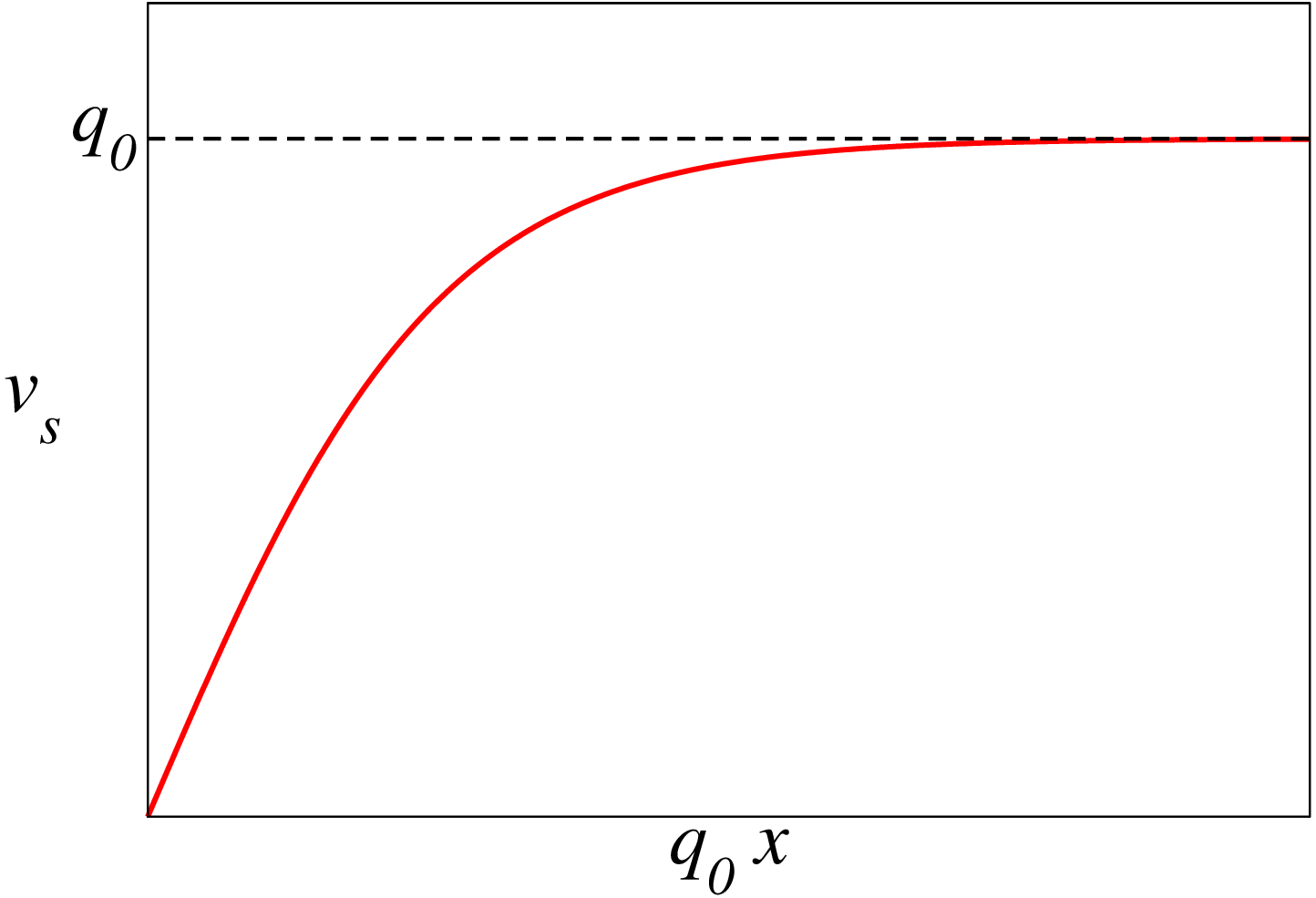}
    \caption[]{The order parameter phase texture near the boundary, see Eq. (\ref{half-DW}), $q_0$ is the FF modulation wavevector in the bulk.}
  	\label{fig: order parameter}
\end{figure}

We have to find a solution, $\Delta(x)$ and $v_s(x)$, of two coupled nonlinear differential equations, Eq. (\ref{re-part-GL-eq}) and $I=0$, where $I$ is given by Eq. (\ref{conserved-I}), subject to the 
boundary conditions 
$$
  \left.\nabla_x\Delta\right|_{x=0}=0,\quad \left.v_s\right|_{x=0}=0,
$$ 
which follow from Eq. (\ref{zero-gradient}). We seek an approximate solution with $\Delta(x)=\Delta_0$ at all $x>0$. Then, the condition $I=0$ takes the following form:
$$
  \frac{d^2v_s}{dx^2} + \frac{|K|}{(1+\varepsilon)\tilde K}v_s - 2v_s^3 = 0,
$$
whose solution vanishing at the boundary is
\begin{equation}
\label{half-DW}
  v_s(x) = q_0 \tanh(q_0x),
\end{equation}
where $q_0$ is given by Eq. (\ref{FF-bulk-wavevector}). This distribution of the order parameter, shown in Fig. \ref{fig: order parameter}, corresponds to one half of the superconducting domain wall found previously 
in Ref. \onlinecite{ST17}. We see that, in order to satisfy the boundary conditions, the phase gradient of the order parameter is strongly deformed near the surface. 
In the bulk of the sample, at $x\gg q_0^{-1}$, we have $v_s\to q_0$, so that a pure FF state is restored. Substituting the expression (\ref{half-DW}) in Eq. (\ref{re-part-GL-eq}), 
one can show that the constant-magnitude approximation is valid if $K^2/\beta\tilde K\Delta_0^2\ll 1$.

\subsection{Magnetic field screening}
\label{sec: B-vs-x}

The magnetic response of the FF state in a half-infinite sample can be found by 
substituting $\psi(x)=\Delta_0e^{i\theta(x)}$, where $\nabla_x\theta$ is given by Eq. (\ref{half-DW}), into the supercurrent (\ref{j_1-anisotropic}). In this way we obtain: $j_{s,y}^{(1)}(x)=-Q(x)A(x)$, where
\begin{equation}
\label{kernel-Q}
  Q(x)=-\frac{8e^2}{\hbar^2c}\frac{|K|\Delta_0^2}{1+\varepsilon} \left[\frac{1}{\cosh^2(q_0x)}+\varepsilon\right].
\end{equation}
The Maxwell equation (\ref{Maxwell-eq}) takes the form
\begin{equation}
\label{Meissner-eq}
  \frac{d^2A}{dx^2}=-\frac{32\pi e^2}{\hbar^2c^2}\frac{|K|\Delta_0^2}{1+\varepsilon} \left[\frac{1}{\cosh^2(q_0x)}+\varepsilon\right] A,
\end{equation}
which has to be solved subject to the matching condition at the surface:
\begin{equation}
\label{bc-for-A}
  \left.\frac{dA}{dx}\right|_{x=0}=H_z.
\end{equation}
Note that in the isotropic case ($\varepsilon=0$), we have $Q(x\to\infty)\to 0$, \textit{i.e.} the transverse current response kernel vanishes in the bulk, in agreement with the discussion in Sec. \ref{sec: GL}. 

By making a change of variables $\xi=\tanh(q_0x)$, Eq. (\ref{Meissner-eq}) can be brought to the following form:
\begin{equation}
\label{ALE}
  \frac{d}{d\xi}\left[(1-\xi^2)\frac{dA}{d\xi}\right]+\left[\nu(\nu+1)-\frac{\mu^2}{1-\xi^2}\right]A=0,
\end{equation}
where 
\begin{eqnarray*}
  \nu=\frac{1}{2}\sqrt{1+\frac{256\pi e^2}{\hbar^2c^2}\tilde K\Delta_0^2}-\frac{1}{2},\quad \mu^2=\frac{64\pi e^2}{\hbar^2c^2}\tilde K\Delta_0^2|\varepsilon|.
\end{eqnarray*}
The differential equation (\ref{ALE}) is known as the associated Legendre equation. Its general solution appropriate for $-1<\xi<1$ is given by 
\begin{equation}
\label{A-general}
  A(\xi)=c_1{\mathrm P}_\nu^{\mu}(\xi)+c_2{\mathrm P}_\nu^{-\mu}(\xi),
\end{equation}
where ${\mathrm P}_\nu^{\mu}(\xi)$ is the Ferrers function of the first kind.\cite{Ferrers-functions} For a solution which is nonsingular in the whole interval $-1<\xi<1$, 
the parameters $\mu$ (known as the order) and $\nu$ (the degree)
have to be non-negative integers and the Ferrers functions become the associated Legendre polynomials. However, since in our case the variable $\xi$ ranges only between $0$ (which corresponds to $x=0$) and $1$ 
(which corresponds to $x\to+\infty$), 
$\mu$ and $\nu$ can be any real positive numbers, related to each other through $\mu^2=|\varepsilon|\nu(\nu + 1)$. In the GL regime, $\Delta_0$ is small, which justifies the assumption that 
$0\leq\mu,\nu\lesssim 1$ for the physically relevant values of the parameters.  

The magnetic induction has the form $B(\xi)=q_0(1-\xi^2)(dA/d\xi)$ and the physical solution is obtained by requiring that the induction does not diverge at $x\to+\infty$. Using the asymptotic formula\cite{Ferrers-functions}
\begin{equation}
\label{P-asymp}
  \left.{\mathrm P}_\nu^\mu(\xi)\right|_{\xi\to 1^-}\sim \frac{1}{\Gamma(1-\mu)}\left(\frac{2}{1-\xi}\right)^{\mu/2},
\end{equation}
where $\Gamma(z)$ is the Gamma function, it is easy to show that $c_1=0$. 
The remaining coefficient $c_2$ is found from the boundary condition $B(\xi=0)=H_z$, see Eq. (\ref{bc-for-A}).
Since the derivative of the Ferrers function is given by
$$
  (1-\xi^2)\frac{d{\mathrm P}_\nu^\mu(\xi)}{d\xi}=(\mu-\nu-1){\mathrm P}_{\nu+1}^\mu(\xi)+(\nu+1)\xi{\mathrm P}_\nu^\mu(\xi),
$$
we obtain:
\begin{equation}
\label{A-final}
  A(\xi)=-\frac{H_z}{q_0(\mu+\nu+1){\mathrm P}_{\nu+1}^{-\mu}(0)}{\mathrm P}_\nu^{-\mu}(\xi)
\end{equation}
and 
\begin{equation}
\label{B-final}
  B(\xi)=\frac{H_z}{{\mathrm P}_{\nu+1}^{-\mu}(0)}\left[{\mathrm P}_{\nu+1}^{-\mu}(\xi)-\frac{\nu+1}{\mu+\nu+1}\xi{\mathrm P}_\nu^{-\mu}(\xi)\right].
\end{equation}
We can also calculate the supercurrent $j_s(x)\equiv j_{s,y}^{(1)}(x)=-Q(x)A(x)$. It follows from Eqs. (\ref{kernel-Q}) and (\ref{A-final}) that
\begin{equation}
\label{screening-j}
  j_s(\xi)=-j_0\frac{1}{(\mu+\nu+1){\mathrm P}_{\nu+1}^{-\mu}(0)}(1-|\varepsilon|-\xi^2){\mathrm P}_\nu^{-\mu}(\xi),
\end{equation}
where
$$
  j_0 = \frac{8e^2}{\hbar^2 c}\frac{|K|\Delta_{0}^2}{q_0(1-|\varepsilon|)}H_z.
$$
We see that, in contrast to the usual case, the $x$-dependence of the induction and the supercurrent is not exponential, although an exponential asymptotics is recovered far from the surface, see below.  

The expressions for $A$, $B$, and $j_s$ all contain
\begin{equation}
\label{P-zero}
  {\mathrm P}_{\nu+1}^{-\mu}(0)=\frac{2^{-\mu}\sqrt{\pi}}{\Gamma\left(\frac{\mu+\nu+3}{2}\right)\Gamma\left(\frac{\mu-\nu}{2}\right)},
\end{equation}
see Ref. \onlinecite{Ferrers-functions}, which reveals an interesting feature of the magnetic response of the FF state. At $\mu-\nu=-2n$, where $n$ is a non-negative integer, Eq. (\ref{P-zero}) passes through zero, 
indicating a singularity in $A(\xi)$ and a change in the behavior of $B(\xi)$. Focusing, as explained above, 
on sufficiently small values of $\mu$ and $\nu$ ($\mu,\nu\lesssim 1$), we can put $n=0$, therefore the singularity occurs when $\mu=\nu$, \textit{i.e.} at $\nu=\nu_c$, where
\begin{equation}
	\nu_{c} = \frac{|\varepsilon|}{1 - |\varepsilon|}.
    \label{nu-critical}
\end{equation}
At $\nu=\nu_c$, the linear response approximation fails and the singularity is expected to be cut off by the higher order terms in the supercurrent 
expansion in powers of $A$. Since the nonlinear effects are relevant only at some exceptional values of the parameters, we leave their investigation outside the scope of the present work and consider only $\nu\neq\nu_c$.

\begin{figure}
    \centering
  	\includegraphics[width=0.60\linewidth]{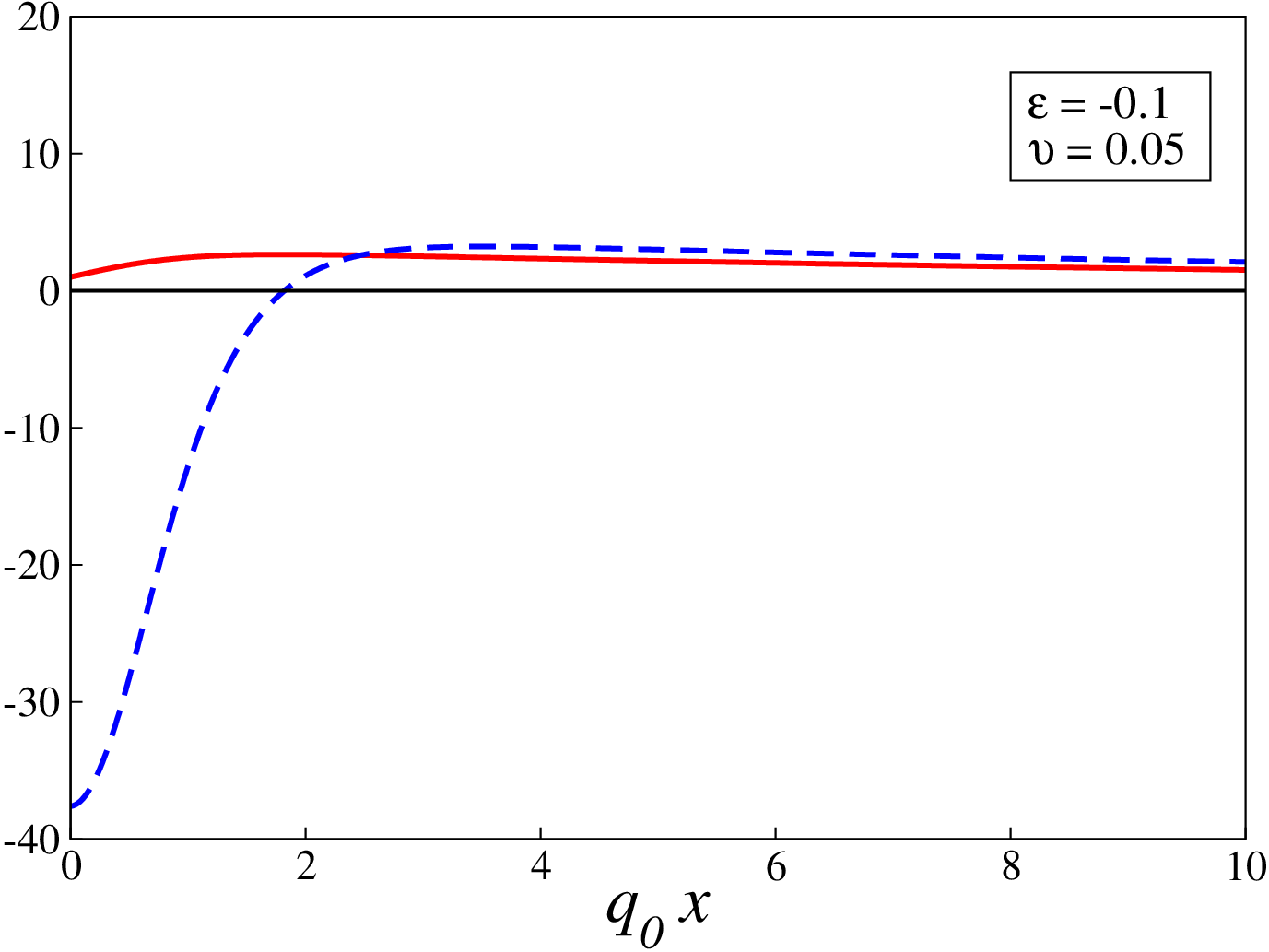}
    \caption[]{The Meissner effect in a weakly anisotropic case, in the anti-screening regime ($\nu<\nu_c\simeq 0.11$). The solid red line is the dimensionless magnetic induction $B(x)/H_z$ and the dashed blue line is 
               the dimensionless supercurrent density $j_s(x)/j_0$.}
  	\label{fig: Meissner-1}
\end{figure}

\begin{figure}
    \centering
  	\includegraphics[width=0.60\linewidth]{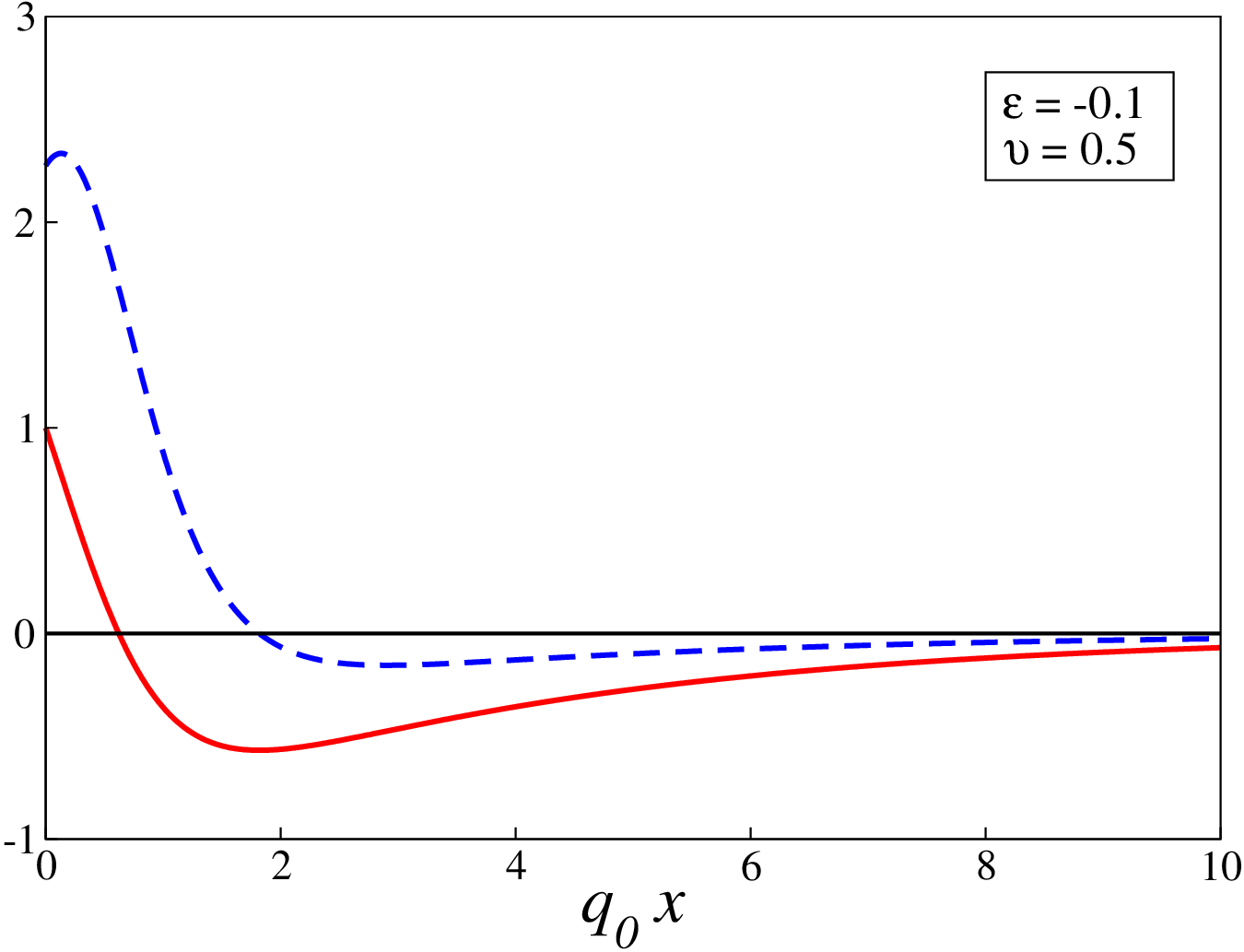}
    \caption[]{The Meissner effect in a weakly anisotropic case, in the over-screening regime ($\nu>\nu_c\simeq 0.11$). The solid red line is the dimensionless magnetic induction $B(x)/H_z$ and the dashed blue line is 
               the dimensionless supercurrent density $j_s(x)/j_0$.}
  	\label{fig: Meissner-2}
\end{figure}

\begin{figure}
    \centering
  	\includegraphics[width=0.60\linewidth]{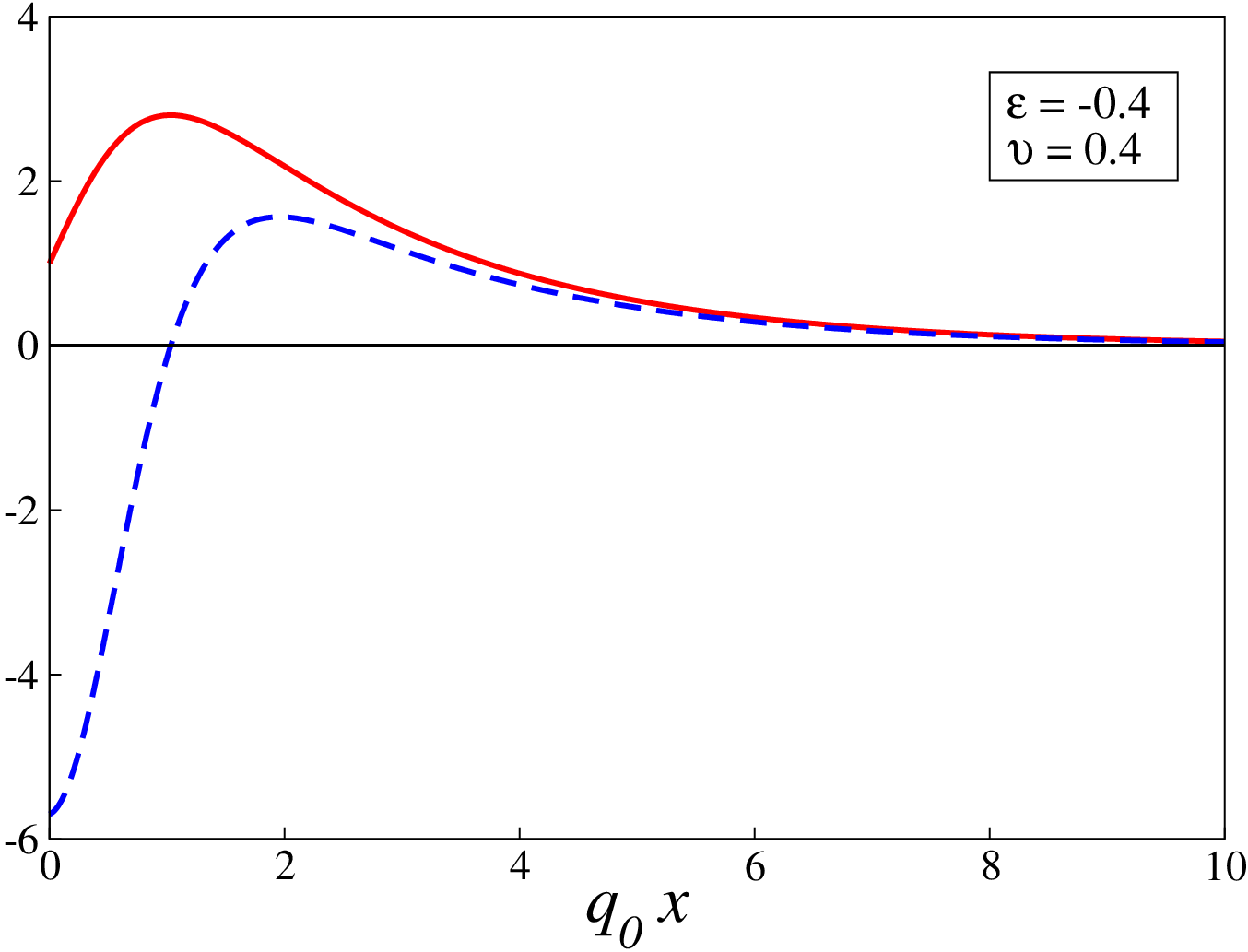}
    \caption[]{The Meissner effect in a strongly anisotropic case, in the anti-screening regime ($\nu<\nu_c\simeq 0.67$). The solid red line is the dimensionless magnetic induction $B(x)/H_z$ and the dashed blue line is 
               the dimensionless supercurrent density $j_s(x)/j_0$.}
  	\label{fig: Meissner-3}
\end{figure}

\begin{figure}
    \centering
  	\includegraphics[width=0.60\linewidth]{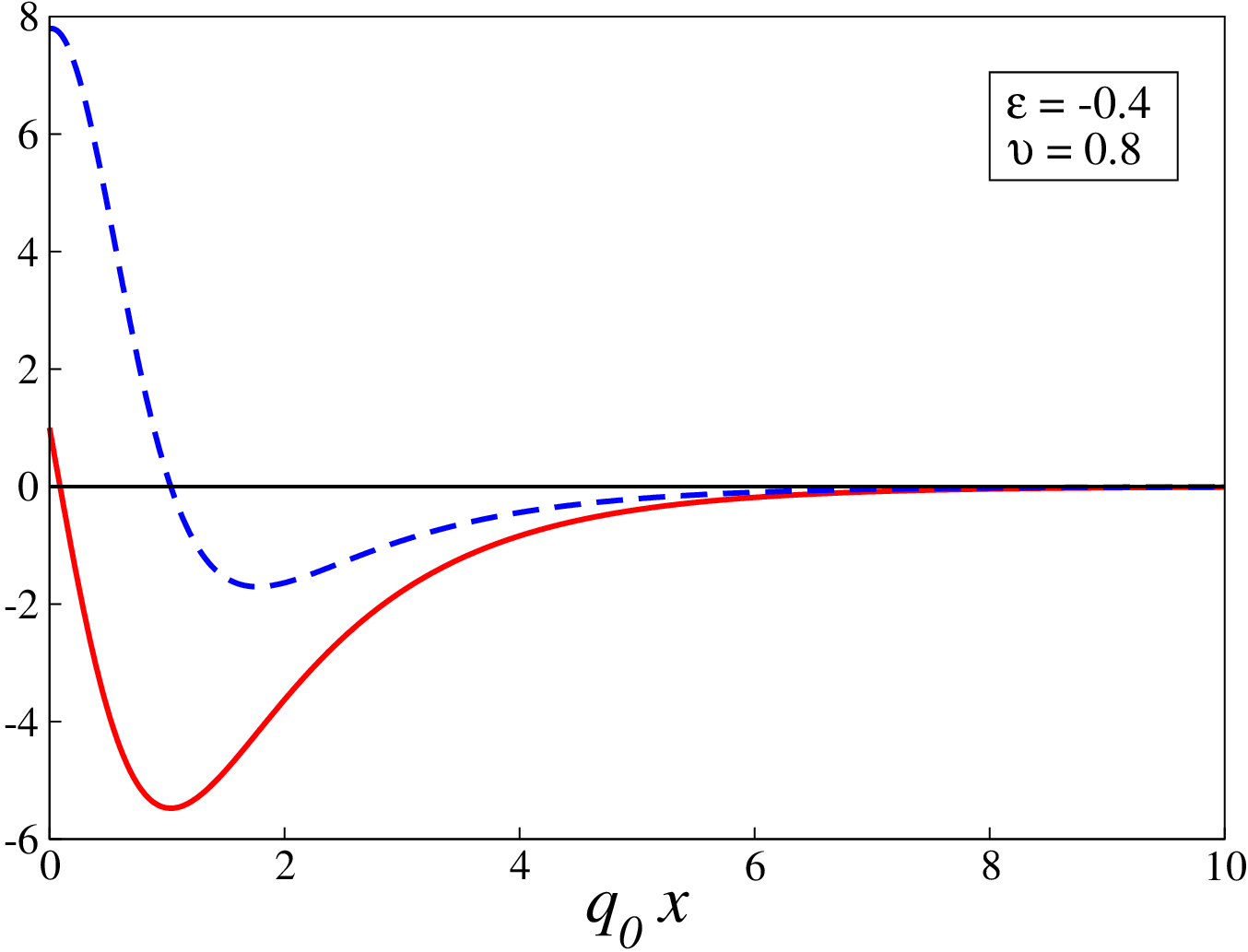}
    \caption[]{The Meissner effect in a strongly anisotropic case, in the over-screening regime ($\nu>\nu_c\simeq 0.67$). The solid red line is the dimensionless magnetic induction $B(x)/H_z$ and the dashed blue line is 
               the dimensionless supercurrent density $j_s(x)/j_0$.}
  	\label{fig: Meissner-4}
\end{figure}

The magnetic field distribution inside the sample turns out to be qualitatively different for $\nu<\nu_c$ and $\nu>\nu_c$. 
In Figs. \ref{fig: Meissner-1}, \ref{fig: Meissner-2}, \ref{fig: Meissner-3}, and \ref{fig: Meissner-4}, this is illustrated by plotting the induction $B(x)$, Eq. (\ref{B-final}), as well as the 
screening supercurrent $j_s(x)$, Eq. (\ref{screening-j}), in the cases of a weak ($\varepsilon=-0.1$) and strong ($\varepsilon=-0.4$) anisotropy. We see that, although the Meissner effect is present in the FF superconductor, 
\textit{i.e.} the magnetic field is expelled from the bulk, it looks very different
from the usual case. The most prominent novel features are the field enhancement (``anti-screening'') and the field inversion (``over-screening''). 
At $\nu<\nu_c$, the magnetic induction initially increases near the surface, then has a maximum, and eventually
decreases to zero at $x\to\infty$. In contrast, at $\nu>\nu_c$, the screening is so strong that the induction changes sign before decaying to zero at infinity. Since $\nu$ depends on temperature through $\Delta_0$, 
the transition between the two regimes should occur at some temperature $\tilde T$ below $T_c$, which can be estimated as follows: 
$$
  \tilde T(\varepsilon)=T_c-\frac{\hbar^2c^2\beta}{64\pi e^2a\tilde K}\frac{|\varepsilon|}{(1-|\varepsilon|)^2},
$$
if the nonlinear effects are neglected. The magnetic field is anti-screened at $\tilde T<T<T_c$ and over-screened at $T<\tilde T$. 

These unusual features of the Meissner response can be verified analytically by calculating the initial slope of $B(x)$ at the surface and also its asymptotics at $x\to\infty$. 
Substituting Eq. (\ref{screening-j}) into the Maxwell equation $\nabla_xB=-4\pi j_s/c$ and using the expression (\ref{P-zero}), we obtain:
\begin{equation}
\label{B-slope}
  \left. \frac{d B}{d x} \right|_{x = 0}=\frac{32\pi e^2}{\hbar^2c^2}\frac{|K|\Delta_0^2H_z}{(\mu+\nu+1)q_0}
	    \frac{\Gamma\left(\frac{\mu+\nu+3}{2}\right)\Gamma\left(\frac{\mu-\nu}{2}\right)}{\Gamma\left(\frac{\mu+\nu+2}{2}\right)\Gamma\left(\frac{\mu-\nu+1}{2}\right)}.
\end{equation}
On the other hand, Eqs. (\ref{B-final}) and (\ref{P-asymp}) yield
\begin{equation}
\label{B-asymp-1}
  B(\xi\to 1^-)\sim \frac{H_z}{{\mathrm P}_{\nu+1}^{-\mu}(0)}(1-\xi)^{\mu/2}\left.\left[\mu+(\nu+1)(1-\xi)\right]\right|_{\xi\to 1^-}. 
\end{equation}
In the general anisotropic case we have $\mu\neq 0$, therefore
\begin{equation}
\label{B-asymp-2}
  \frac{B(x\to\infty)}{H_z}\sim\mu\Gamma\left(\frac{\mu-\nu}{2}\right)e^{-\mu q_0x},
\end{equation}
where we again used Eq. (\ref{P-zero}). Note that the same Gamma function $\Gamma[(\mu-\nu)/2]$ appears in both the small and large $x$ asymptotics of the magnetic induction. At $\nu<\nu_c$, we have $\mu>\nu$, therefore
$B'(0)>0$ and $B(+\infty)=0^+$, \textit{i.e.} the field is initially enhanced and then decays without changing sign. At $\nu>\nu_c$, we have $\mu<\nu$, therefore $B'(0)<0$ and $B(+\infty)=0^-$, \textit{i.e.} the field decreases so fast that it reverses
its direction at some point.

To examine the transition from a weakly anisotropic to the fully isotropic case, one can keep the parameter $\nu$ fixed while reducing the value of $|\varepsilon|$. According to Eq. (\ref{nu-critical}), 
the system will eventually enter the over-screening regime with $\nu>\nu_c$. One can show that as $\varepsilon\to 0^-$, the point at which the magnetic induction changes sign moves to infinity, 
therefore no field inversion is expected to occur in the isotropic limit. This is confirmed by an explicit calculation in the next subsection.

\subsection{Isotropic case}
\label{sec: Meissner-isotropic}

The isotropic case is recovered from Eqs. (\ref{B-final}) and (\ref{screening-j}) by taking the limit $\mu\to 0$, in which the Ferrers functions become the Legendre functions: 
$$
  {\mathrm P}_\nu^0(\xi)={\mathrm P}_\nu(\xi)=F\left(\nu+1,-\nu;1;\frac{1-\xi}{2}\right),
$$
where $F(a,b;c;z)$ is the hypergeometric function (Ref. \onlinecite{Ferrers-functions}). The magnetic induction and the screening supercurrent take the following form:
\begin{equation}
\label{B-isotropic}
  B(\xi)=H_z\frac{\Gamma\left(\frac{\nu+3}{2}\right)\Gamma\left(-\frac{\nu}{2}\right)}{\sqrt{\pi}}\left[{\mathrm P}_{\nu+1}(\xi)-\xi{\mathrm P}_\nu(\xi)\right]
\end{equation}
and
\begin{equation}
\label{j_s-isotropic}
  j_s(\xi)=-j_0\frac{\Gamma\left(\frac{\nu+3}{2}\right)\Gamma\left(-\frac{\nu}{2}\right)}{\sqrt{\pi}(\nu+1)}(1-\xi^2){\mathrm P}_\nu(\xi),
\end{equation}
respectively. 

\begin{figure}
    \centering
  	\includegraphics[width=0.60\linewidth]{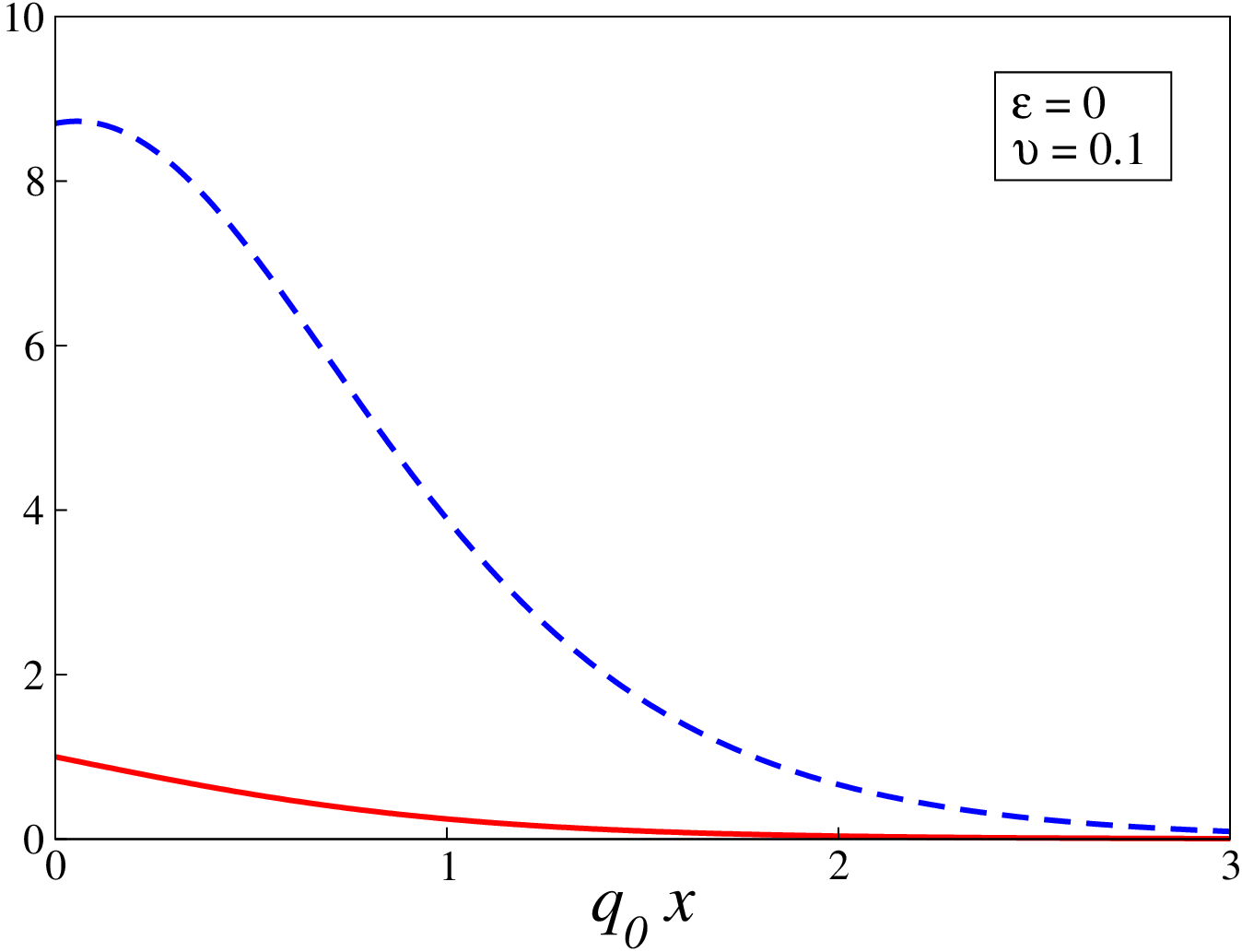}
    \caption[]{The Meissner effect in the isotropic case, for a small $\nu$. The solid red line is the dimensionless magnetic induction $B(x)/H_z$ and the dashed blue line is 
               the dimensionless supercurrent density $j_s(x)/j_0$.}
  	\label{fig: Meissner-5}
\end{figure}

\begin{figure}
    \centering
  	\includegraphics[width=0.60\linewidth]{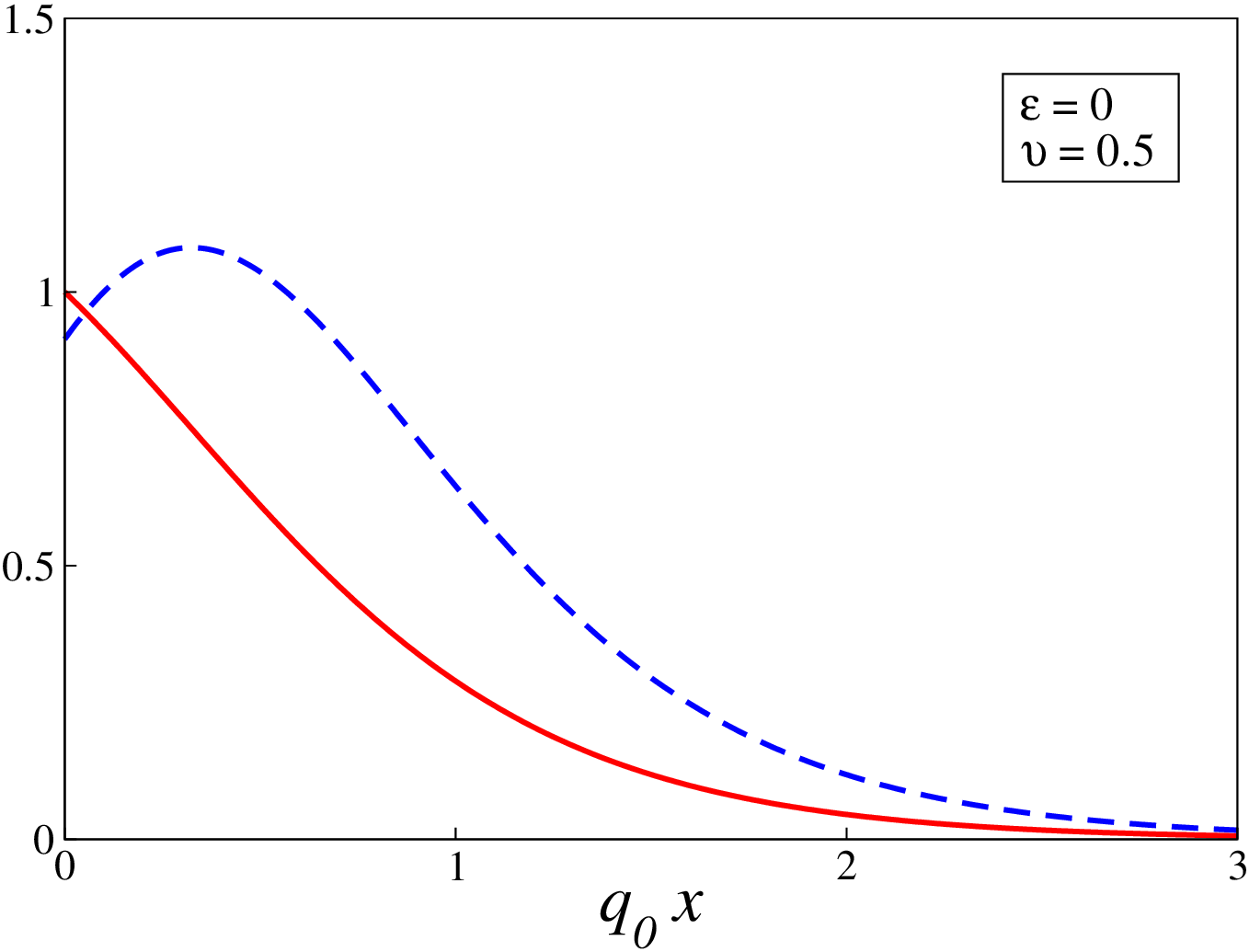}
    \caption[]{The Meissner effect in the isotropic case, for a large $\nu$. The solid red line is the dimensionless magnetic induction $B(x)/H_z$ and the dashed blue line is 
               the dimensionless supercurrent density $j_s(x)/j_0$.}
  	\label{fig: Meissner-6}
\end{figure}

It follows from Eq. (\ref{B-isotropic}) that the magnetic field is expelled from the superconducting bulk, with an exponential asymptotics far from the surface: $B(\xi\to 1^-)\sim H_z(1-\xi)$, see also Eq. (\ref{B-asymp-1}), 
therefore  
\begin{equation}
\label{B-asymp-isotropic}
  \frac{B(x\to\infty)}{H_z}\sim e^{-2 q_0x}.
\end{equation}
Similarly, the screening current decays exponentially, as $j_s(x\to\infty)/j_0\sim e^{-2 q_0x}$. We have plotted $B(x)$ and $j_s(x)$ in Figs. \ref{fig: Meissner-5} and \ref{fig: Meissner-6}. 
We see that the isotropic FF state does exhibit the Meissner effect, which originates from the order parameter deformation near the surface. 
The decay of both the magnetic induction and the supercurrent can be characterized by an effective penetration depth $\lambda_{FF}=(2q_0)^{-1}$, 
which is of the order of the FF modulation wavelength.

\section{Surface critical field}
\label{sec: Hc3}

It is well known that if a half-infinite non-FFLO superconductor is placed in a magnetic field parallel to the surface, then superconductivity first appears in the field
equal to the surface critical field $H_{c3}$ (Ref. \onlinecite{DeGennes-book}). In this section, we calculate the surface critical field in the FFLO case. 
We only consider the isotropic system ($\varepsilon=0$), in which case an exact analytical solution is possible.    

Assuming that the phase transition is of the second order, the surface superconducting instability can be studied by solving the linearized GL equation, supplemented by 
the boundary conditions (\ref{zero-gradient}) and (\ref{zero-normal-current}), with an additional requirement that the order parameter is localized near the surface, \textit{i.e.} $\psi|_{x\to\infty}=0$. 
From Eq. (\ref{GL-energy-density-anisotropic}) we obtain:
\begin{equation}
\label{GL-eq-linear}
  \alpha\psi+|K|(D_x^2+D_y^2)\psi+\tilde K(D_x^2+D_y^2)^2\psi=0.
\end{equation}
Using the Landau gauge, $\bA=H_zx\hby$ (recall that the ``orbital'' pair breaking is due to the component of $\bm{H}$ perpendicular to the basal plane), the solution can be written as 
\begin{equation}
\label{separation-of-variables}
  \psi(\br)=e^{ik_yy}f(x),
\end{equation}
where the function $f$ satisfies the following equation:
\begin{equation}
\label{eq-for-f}
  \alpha f-|K|\hat Lf+\tilde K\hat L^2f=0. 
\end{equation}
The notations here are as follows:
$$
  \hat L=-\frac{d^2}{dx^2}+h^2(x-x_0)^2,
$$
$h=2\pi H_z/\Phi_0$, $x_0=-k_y/h$, and $\Phi_0=\pi\hbar c/e$ is the magnetic flux quantum. Without loss of generality, we assume that $H_z>0$.
From Eq. (\ref{eq-for-f}) we obtain that the critical temperature as a function of the applied ``orbital'' field is given by
\begin{equation}
\label{T_c-vs-H}
  T_c(H_z)=T_c(0)-\frac{1}{a}\Lambda_{min}(H_z),
\end{equation}
where $T_c(0)=T_{c,0}+K^2/4a\tilde K$ is the critical temperature of the zero-field FFLO transition, $\Lambda_{min}$ is the lowest eigenvalue of the operator
$$
  \hat\Lambda=-|K|\hat L+\tilde K\hat L^2+\frac{K^2}{4\tilde K}=\tilde K(\hat L-q_0^2)^2, 
$$
and $q_0=\sqrt{|K|/2\tilde K}$.

\begin{figure}
    \centering
  	\includegraphics[width=0.60\linewidth]{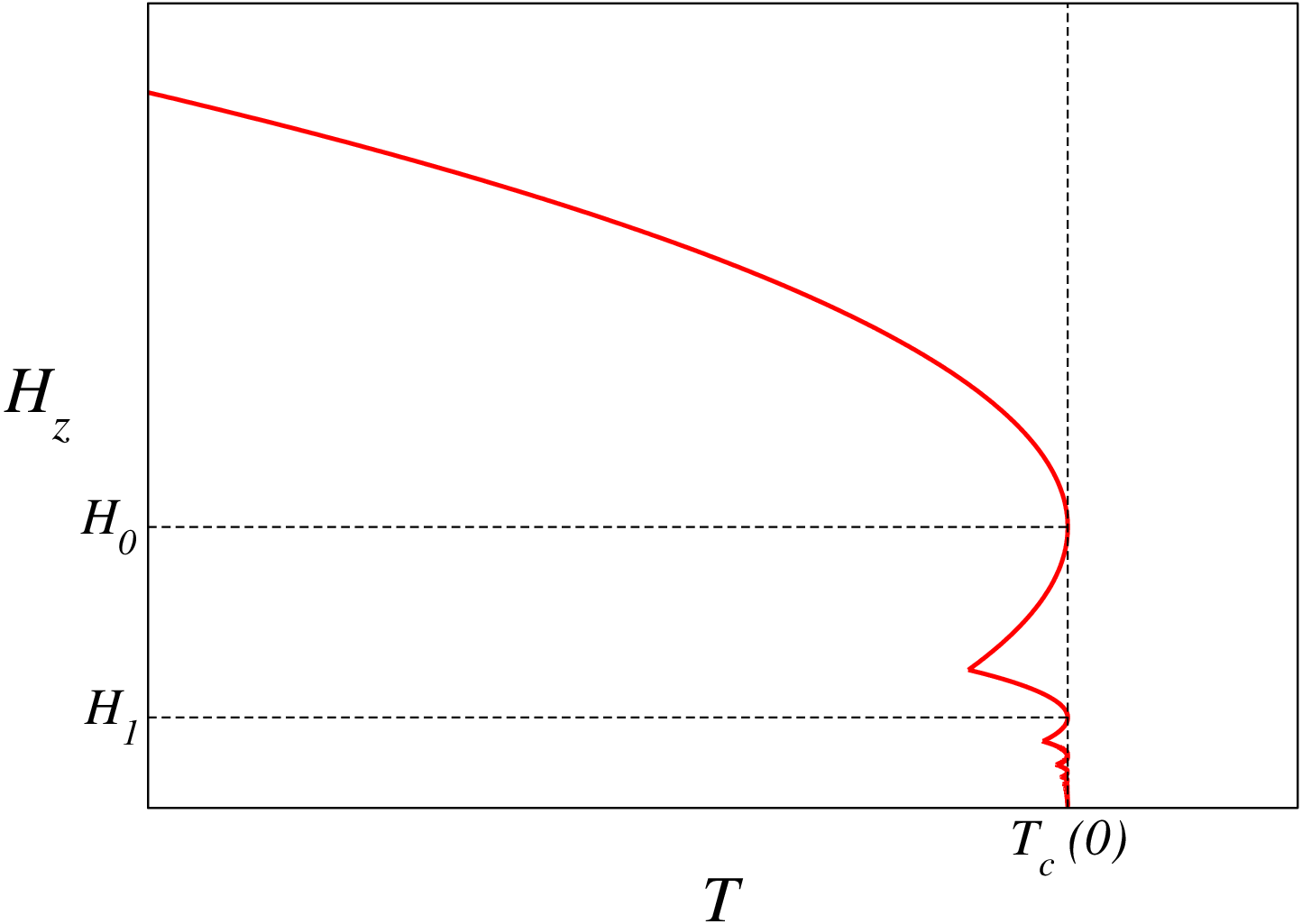}
    \caption[]{The upper critical field $H_{c2}(T)$ of an isotropic quasi-2D FFLO superconductor, after Ref. \onlinecite{BK84}.}
  	\label{fig: Hc2}
\end{figure}

First, let us calculate the upper critical field $H_{c2}$, by solving Eq. (\ref{eq-for-f}) in an infinite sample. Due to translational invariance one can set $x_0=0$. The common eigenfunctions of the operators 
$\hat L$ and $\hat\Lambda$ are the harmonic oscillator wave functions $f_n(x)\propto e^{-hx^2/2}H_n(\sqrt{h}x)$,
where $n=0,1,2,...$ and $H_n(z)$ are the Hermite polynomials. The eigenvalues of $\hat L$ are given by the Landau levels $(2n+1)h$, and those of $\hat\Lambda$ are
$\Lambda_n=\tilde K\left[(2n+1)h-q_0^2\right]^2$. Therefore, for the critical temperature in the bulk we obtain:
\begin{equation}
\label{T_c-bulk}
  T_c(H_z)=T_c(0)-\left(\frac{2\pi}{\Phi_0}\right)^2\frac{\tilde K}{a}\min_n\left[(2n+1)H_z-\frac{\Phi_0|K|}{4\pi\tilde K}\right]^2.
\end{equation}
As the out-of-plane magnetic field changes, the ground state of $\hat\Lambda$ switches between different Landau levels and the critical temperature shows the Little-Parks oscillations, see 
Ref. \onlinecite{BK84} and Fig. \ref{fig: Hc2}. The critical temperature reaches its maximum value, equal to $T_c(0)$, at $H_z=H_n$, where
\begin{equation}
\label{H_n}
  H_n=\frac{1}{2n+1}\frac{\Phi_0|K|}{4\pi\tilde K}.
\end{equation}
Switching from the $n$th Landau level to $(n+1)$th Landau level occurs at $H_z=\tilde H_n=\Phi_0|K|/8(n+1)\pi\tilde K$, which satisfies $H_{n+1}<\tilde H_n<H_n$.   

\begin{figure}
    \centering
  	\includegraphics[width=0.60\linewidth]{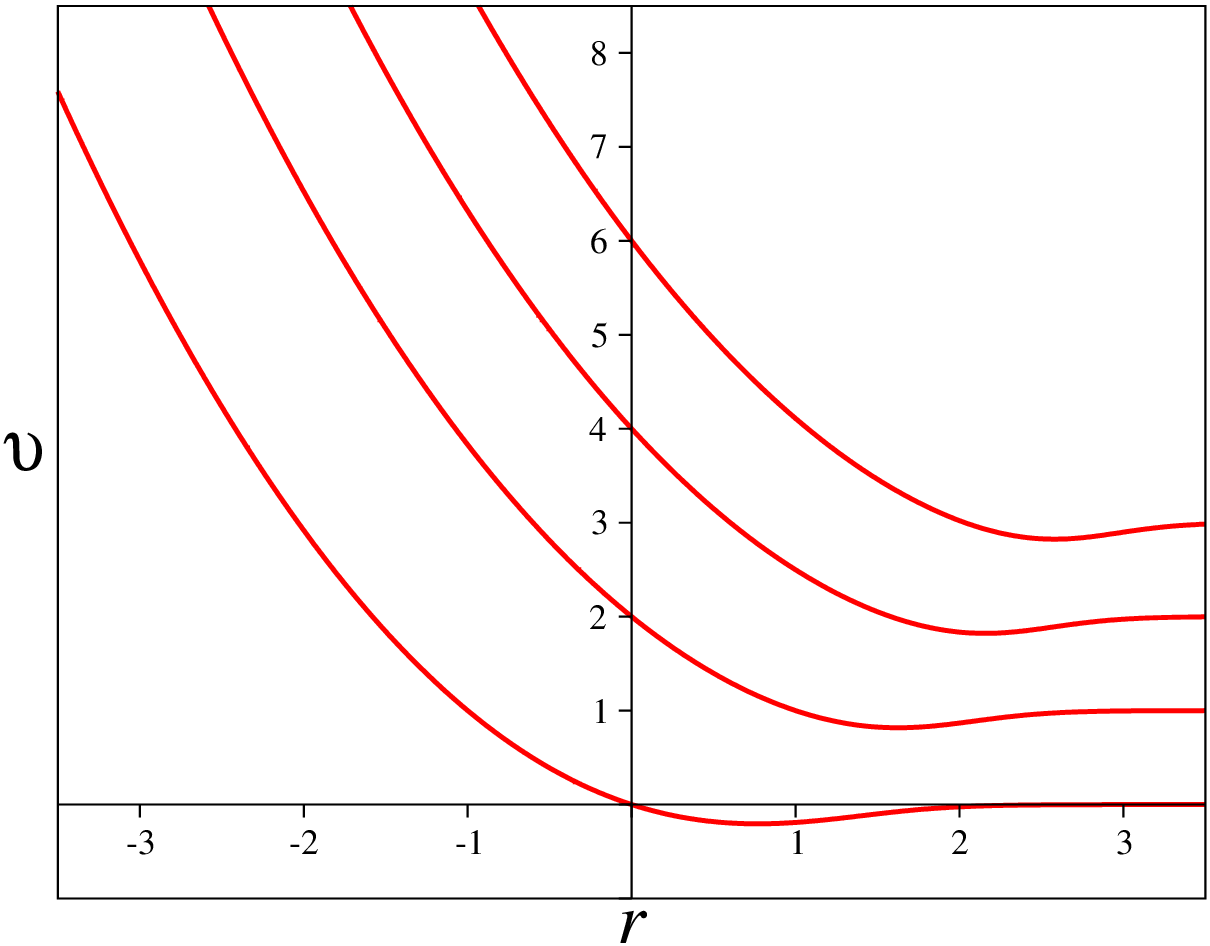}
    \caption[]{The first four solutions of Eq. (\ref{nu-vs-r}), as functions of $r=\sqrt{h}x_0$.}
  	\label{fig: Hermite}
\end{figure}

Now let us look at the surface superconductivity in a half-infinite sample. The boundary condition (\ref{zero-normal-current}) is satisfied automatically for the order parameter (\ref{separation-of-variables}) 
with a real $f$. We further require that 
\begin{equation}
\label{f-bc-1}
  f'(0)=0,
\end{equation}
according to Eq. (\ref{zero-gradient}), and also that $f(\infty)=0$, corresponding to a superconductiving nucleus localized near the surface. 
The eigenfunctions of the operator $\hat L$ are given by 
\begin{equation}
\label{L-eigenfunctions}
  f_\nu(x)\propto e^{-h(x-x_0)^2/2}H_\nu[\sqrt{h}(x-x_0)],
\end{equation}
where $H_\nu(z)$ is the Hermite function,\cite{NikUv-book} and $\hat Lf_\nu=(2\nu+1)hf_\nu$. In contrast to the bulk case, we require that the eigenfunctions vanish only at $x\to+\infty$, 
therefore the index $\nu$ does not have to be integer (for non-negative integer values of $\nu$, the Hermite functions become the Hermite polynomials). Substituting 
Eq. (\ref{L-eigenfunctions}) into the boundary condition (\ref{f-bc-1}) and using the property $H_\nu'(z)=2\nu H_{\nu-1}(z)$, we obtain the following equation for $\nu$:
\begin{equation}
\label{nu-vs-r}
  2\nu H_{\nu-1}(-r)=-rH_{\nu}(-r),
\end{equation}
where $r=\sqrt{h}x_0$. The roots of this last equation determine the eigenvalues of the operator $\hat\Lambda$ at given $r$:
\begin{equation}
\label{lambda_nu}
  \Lambda_\nu=\tilde K\left[(2\nu+1)h-q_0^2\right]^2.
\end{equation}
The lowest four solutions of Eq. (\ref{nu-vs-r}) are plotted in Fig. \ref{fig: Hermite}. Note that at $r\to\infty$, the bulk solution with $\nu=n=0,1,2,...$ is recovered.
One can show that the absolute minimum of $\nu(r)$ is achieved at $r=r^*\simeq 0.768$ and given by 
$$
  \nu^*\simeq -0.205.
$$ 
From this we obtain the well-known expression for the surface critical field in the non-FFLO case: $H_{c3}=H_{c2}/(2\nu^*+1)\simeq 1.695H_{c2}$ (Ref. \onlinecite{DeGennes-book}). 

According to Eq. (\ref{T_c-vs-H}), we have to minimize the eigenvalues (\ref{lambda_nu}) with respect to $r$, subject to the constraint that $\nu>\nu^*$. 
Since
$$
  \frac{\partial\Lambda_\nu}{\partial r}\propto \left[(2\nu+1)h-q_0^2\right]\frac{d\nu}{dr},
$$
the minimum of $\Lambda_\nu$ corresponds to either 
\begin{equation}
\label{nu-low-h}
  \nu=\frac{1}{2}\left(\frac{q_0^2}{h}-1\right),
\end{equation}
or $d\nu/dr=0$, \textit{i.e.} 
\begin{equation}
\label{nu-high-h}
  \nu=\nu^*.
\end{equation}
The first possibility leads to $\Lambda_{min}=0$, therefore the critical temperature is not affected by the field: $T_c(H_z)=T_c(0)$. However, this can only be realized if $\nu>\nu^*$, 
\textit{i.e.} at $h<h^*=q_0^2/(2\nu^*+1)$. At higher fields, we have the second possibility, with $\Lambda_{min}=\tilde K\left[(2\nu^*+1)h-q_0^2\right]^2$. 

\begin{figure}
    \centering
  	\includegraphics[width=0.60\linewidth]{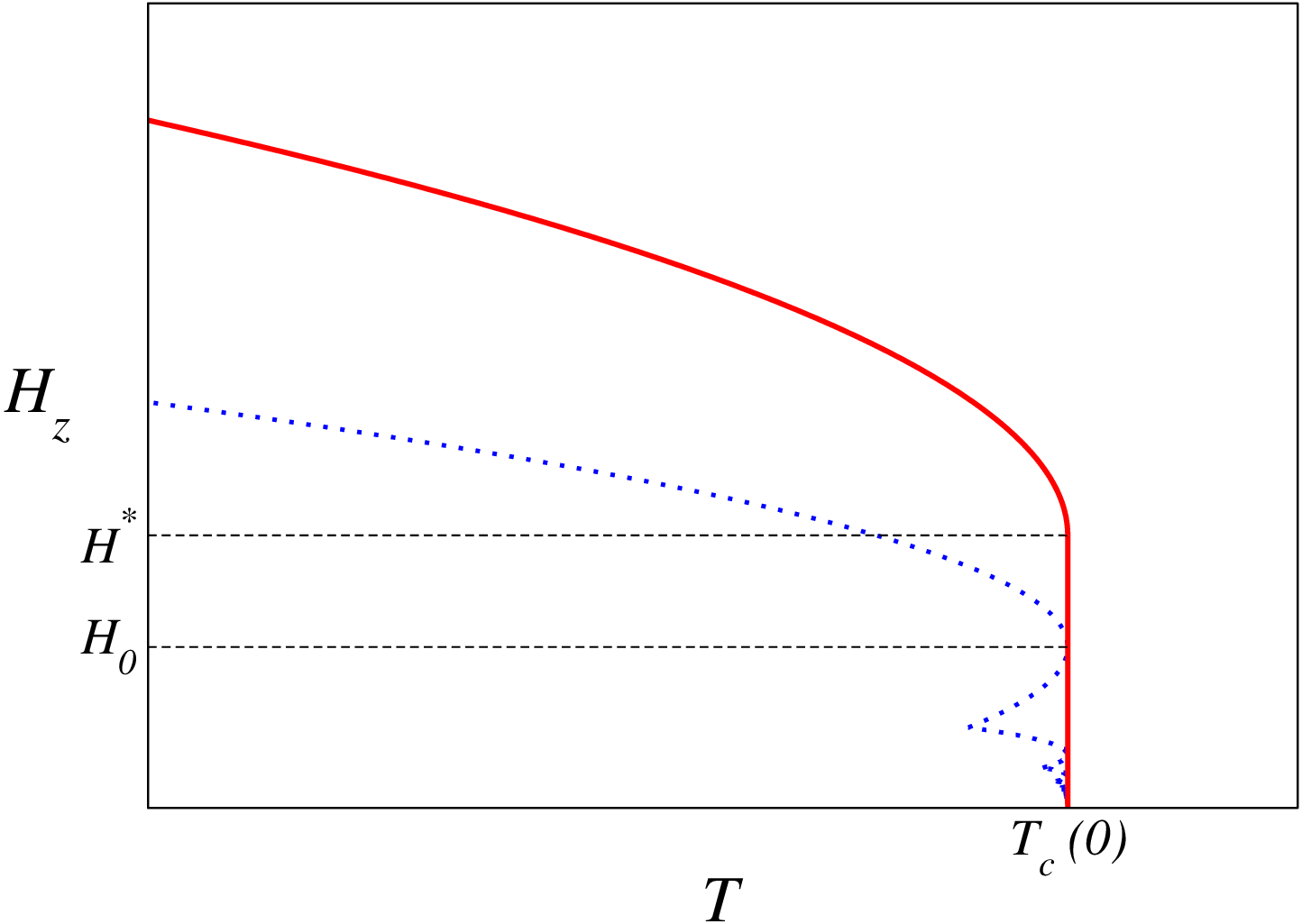}
    \caption[]{The surface critical field $H_{c3}(T)$ of an isotropic quasi-2D FFLO superconductor (the solid red line), with $H^*\simeq 1.695 H_0$. The blue dotted line shows $H_{c2}(T)$ for comparison.}
  	\label{fig: Hc3}
\end{figure}

Collecting everything together, we finally arrive at the following expression for the critical temperature of the FFLO instability near the surface: 
\begin{eqnarray}
\label{H_c3-final}
  && T_c(H_z)=T_c(0),\quad 0\leq H_z<H^*,\nonumber\\ && \\
  && T_c(H_z)=T_c(0)-(2\nu^*+1)^2\left(\frac{2\pi}{\Phi_0}\right)^2\frac{\tilde K}{a}\left(H_z-H^*\right)^2,\quad H_z>H^*,\nonumber
\end{eqnarray}
which is plotted in Fig. \ref{fig: Hc3}. Here 
$$
  H^*=\frac{1}{2\nu^*+1}\frac{\Phi_0|K|}{4\pi\tilde K}\simeq 1.695 H_0,
$$
and $H_0$ is the field below which the bulk critical temperature exhibits the Little-Parks oscillations, see Eq. (\ref{H_n}). 
We would like to mention the study of the FFLO state in a disk geometry in Ref. \onlinecite{SMB10}, whose numerical results are qualitatively consistent with our analytical calculation.  
  
Regarding the shape of the superconducting nucleus near the surface, it is given by Eq. (\ref{L-eigenfunctions}) and characterized by two field-dependent parameters, $\nu$ and $r$. 
For the index $\nu$, we have Eq. (\ref{nu-high-h}) at high fields ($H>H^*$) and 
Eq. (\ref{nu-low-h}) at low fields ($H<H^*$), while $r$ is found by solving Eq. (\ref{nu-vs-r}). There is only one solution $r=r^*$ at high fields, but in the low-field case multiple solutions are possible, as evident
from Fig. \ref{fig: Hermite}. In fact, the number of possible values of $r$ goes to infinity at $H\to 0$, when $\nu\to\infty$ according to Eq. (\ref{nu-low-h}). This means that the order parameter 
is represented by a superposition of several solutions of the form (\ref{L-eigenfunctions}), with the coefficients determined by minimizing the full nonlinear GL free energy. We leave further investigation of the 
order parameter profile to a future work.

\section{Conclusions}
\label{sec: Conclusions}

We studied the ``orbital'' magnetic properties of a half-infinite quasi-2D FFLO superconductor, which can be probed by tilting the applied field out of the basal plane.
We used the modified GL formalism, supplemented by the boundary conditions at a superconductor-insulator interface. Due to the presence of the higher-order gradient terms in the GL equations, the number of 
the boundary conditions increases compared to the usual (non-FFLO) case. We focused on two observable properties: (i) the magnetic field screening (the Meissner effect), which is a result of the linear 
response of a given order parameter configuration to a weak external field, and (ii) the surface critical field $H_{c3}$, in which case the order parameter is small and the linearized modified GL equations 
are solved in an arbitrary magnetic field.

In order to satisfy the boundary conditions, the order parameter near the surface deviates considerably from the single-plane wave FF state, showing a domain wall-like phase texture. 
In the isotropic case, this leads to the generation of the screening currents and the expulsion of magnetic field from the superconductor, despite the vanishing of the transverse 
superfluid density in the bulk. Even in an anisotropic system, when the bulk superfluid density is nonzero, the linear magnetic response turns out to be very different from the usual case. 
Depending on the temperature, the Meissner effect exhibits two different regimes. Near the critical temperature, the field is anti-screened, with the magnetic induction near the surface exceeding
the applied field, passing through a maximum and then decreasing to zero at $x\to\infty$. At lower temperatures, the field is over-screened, with the induction changing sign inside the superconductor, 
before decaying to zero.      

Similar to the usual case, the FFLO superconductivity in an external ``orbital'' magnetic field preferentially nucleates near the surface of the sample. However, the temperature dependence of the surface critical field 
considerably differs both from that in the usual case and also from the FFLO upper critical field in the bulk. In contrast to the latter, $H_{c3}(T)$ does not show any Little-Parks oscillations. At low fields, $H_z<H^*$, 
the surface superconductivity appears at the same temperature as at zero field, \textit{i.e.} $T_c(H_z)=T_c(H_z=0)$, while at higher fields, $H_z>H^*$, the surface superconductivity is suppressed, with 
$T_c(H_z=0)-T_c(H_z)\propto(H_z-H^*)^2$.

\acknowledgments
K. S. is grateful to J.-P. Brison for useful discussions. This work was supported by a Discovery Grant from the Natural Sciences and Engineering Research Council of Canada.

\appendix

\section{Boundary conditions in an FFLO superconductor}
\label{app: boundary conditions}

In this Appendix, we derive the boundary conditions for a modified GL equation using a variational analysis. The GL free energy in a finite sample in the absence of external magnetic field has the form
${\cal F}={\cal F}_{bulk}+{\cal F}_{surface}$, where
\begin{equation}
\label{F-bulk}
  {\cal F}_{bulk}=\int dV\left[\alpha|\psi|^2+\frac{\beta}{2}|\psi|^4 + K|\bnab\psi|^2+\tilde K|\bnab^2\psi|^2\right]
\end{equation}
is the bulk contribution (here we put $\varepsilon=0$ for simplicity). The surface contribution, which phenomenologically describes the modification of the conditions for superconductivity near the surface, 
can be represented as\cite{Book}
\begin{equation}
\label{F-surface}
  {\cal F}_{surface}=\oint dS\;\sigma|\psi|^2,
\end{equation}
where the integration goes over the surface of the sample and $\sigma$ is a real constant. The variation of the total free energy under $\psi^*\to\psi^*+\delta\psi^*$ is given by
\begin{eqnarray*}
  \delta{\cal F}=\int dV\;\delta\psi^*\left(\alpha\psi+\beta|\psi^2|\psi-K\nabla^2\psi+\tilde K\nabla^4\psi\right)\\
		+\oint dS\;\delta\psi^*\left[\sigma\psi+K(\bfn\bnab)\psi-\tilde K(\bfn\bnab)\nabla^2\psi\right]\\
		+\oint dS\;(\bfn\bnab\delta\psi^*)\left(\tilde K\nabla^2\psi\right),
\end{eqnarray*}
where $\bfn$ is the outward normal to the surface. We will also need the expression for the normal component of the supercurrent:
\begin{equation}
\label{j_normal}
  \left.(\bfn\bjs)\right|_S=-\left.\frac{4e}{\hbar} \im \left\{ \psi^*[K(\bfn\bnab)\psi-\tilde K(\bfn\bnab)\nabla^2\psi] + \tilde{K}(\bfn\bnab\psi)^*\nabla^2\psi\right\}\right|_S,
\end{equation}
which follows from Eq. (\ref{j_0}).

Requiring that $\delta{\cal F}=0$ produces, besides the GL equation (\ref{GL-eq-zero-field}) in the bulk, two additional conditions at the surface:
$$
  \left.\delta\psi^*\left[\sigma\psi+K(\bfn\bnab)\psi-\tilde K(\bfn\bnab)\nabla^2\psi\right]\right|_S=0
$$
and
$$
  \left.\delta(\bfn\bnab\psi)^*\left(\tilde K\nabla^2\psi\right)\right|_S=0.
$$
From the first condition we obtain:
\begin{equation}
\label{cond-1}
  \left[K(\bfn\bnab)\psi-\tilde K(\bfn\bnab)\nabla^2\psi\right]_S=-\left.\sigma\psi\right|_S,
\end{equation}
while the second condition is satisfied if one fixes the surface gradient of the order parameter:
\begin{equation}
\label{general-bc-1}
  \left.(\bfn\bnab)\psi\right|_S=0.
\end{equation}
Substituting Eqs. (\ref{cond-1}) and (\ref{general-bc-1}) in the supercurrent (\ref{j_normal}) we have
\begin{equation}
\label{general-bc-2}
  \left.(\bfn\bjs)\right|_S=0.
\end{equation}
Thus the boundary conditions (\ref{zero-gradient}) and (\ref{zero-normal-current}) are reproduced.

\clearpage

\clearpage

\clearpage

\clearpage

\clearpage

\clearpage

\clearpage

\clearpage

\clearpage

\clearpage

\begin{thebibliography}{99}

\bibitem{FF64}
P. Fulde and R. A. Ferrell, Phys. Rev. \textbf{135}, A550 (1964).

\bibitem{LO64}
A. I. Larkin and Yu. N. Ovchinnikov, Zh. Eksp. Teor. Fiz. \textbf{47}, 1136 (1964) [Sov. Phys. JETP \textbf{20}, 762 (1965)].

\bibitem{FFLO-review-07}
Y. Matsuda and H. Shimahara, J. Phys. Soc. Jpn. \textbf{76}, 051005 (2007).

\bibitem{Uji12}
S. Uji, K. Kodama, K. Sugii, T. Terashima, Y. Takahide, N. Kurita, S. Tsuchiya, M. Kimata, A. Kobayashi, B. Zhou, and H. Kobayashi, Phys. Rev. B \textbf{85}, 174530 (2012).

\bibitem{RS09}
L. Radzihovsky and D. E. Sheehy, Rep. Prog. Phys. \textbf{73}, 076501 (2010).

\bibitem{CasNar04}
R. Casalbuoni and G. Nardulli, Rev. Mod. Phys. \textbf{76}, 263 (2004).

\bibitem{DeGennes-book}
P. G. de Gennes, \textit{Superconductivity of Metals and Alloys} (Westview Press, 1999).

\bibitem{BK84}
A. I. Buzdin and M. L. Kuli\'c, J. Low Temp. Phys. \textbf{54}, 203 (1984).

\bibitem{tilted-H}
L. N. Bulaevskii, Zh. Eksp. Teor. Fiz. \textbf{65}, 1278 (1973) [Sov. Phys. JETP \textbf{38}, 634 (1974)];
A. I. Buzdin and J.-P. Brison, Europhys. Lett. \textbf{35}, 707 (1996);
M. Houzet, A. Buzdin, L. Bulaevskii, and M. Maley, Phys. Rev. Lett. \textbf{88}, 227001 (2002).

\bibitem{BK97}
A. I. Buzdin and H. Kachkachi, Phys. Lett. A \textbf{225}, 341 (1997).

\bibitem{AY01}
D. F. Agterberg and K. Yang, J. Phys.: Condens. Matter \textbf{13}, 9259 (2001).

\bibitem{HM06}
M. Houzet and V. P. Mineev, Phys. Rev. B \textbf{74}, 144522 (2006).

\bibitem{BMS07}
A. Buzdin, Y. Matsuda, and T. Shibauchi, Europhys. Lett. \textbf{80}, 67004 (2007).

\bibitem{Tinkham-book}
M. Tinkham, \textit{Introduction to Superconductivity} (McGraw-Hill, New York, 1996).

\bibitem{ST17}
K. V. Samokhin and B. P. Truong, Phys. Rev. B \textbf{96}, 214501 (2017).

\bibitem{zero-stiffness}
V. F. Elesin, Zh. Eksp. Teor. Fiz. \textbf{131}, 938 (2007) [JETP \textbf{104}, 819 (2007)];
K. V. Samokhin, Phys. Rev. B \textbf{83}, 094514 (2011);
L. Radzihovsky, Phys. Rev. A \textbf{84}, 023611 (2011).

\bibitem{Ferrers-functions}
F. W. J. Olver, \textit{Asymptotics and Special Functions} (Academic Press, 1974); \textit{NIST Digital Library of 
Mathematical Functions}, Ch. 14, http://dlmf.nist.gov/, F. W. J. Olver, A. B. Olde Daalhuis, D. W. Lozier, B. I. Schneider, R. F. Boisvert, C. W. Clark, B. R. Miller, and B. V. Saunders, eds.

\bibitem{NikUv-book}
A. F. Nikiforov and V. B. Uvarov, \textit{Special Functions of Mathematical Physics} (Birkh\"auser, 1988).

\bibitem{SMB10}
A. V. Samokhvalov, A. S. Mel'nikov, and A. I. Buzdin, Phys. Rev. B \textbf{82}, 174514 (2010).

\bibitem{Book}
V. P. Mineev and K. V. Samokhin, \textit{Introduction to Unconventional Superconductivity} (Gordon and Breach, London, 1999).

\end{thebibliography}
\end{document}